\documentclass[a4paper,11pt]{article}
\usepackage{jheppub}
\usepackage{graphicx,multirow,subfig}
\usepackage{float}
\usepackage{bm}
\usepackage{braket}
\usepackage{amsmath}
\usepackage{amssymb}
\usepackage{amscd}
\usepackage{latexsym}
\usepackage{slashed}
\usepackage{color}
\usepackage{graphicx}
\usepackage[normalem]{ulem}
\definecolor{orange}{cmyk}{0,0.5,1,0}
\usepackage{verbatim}
\usepackage{rotating}
\usepackage{cancel}
\usepackage{caption}
\usepackage[utf8]{inputenc}

\hypersetup{
   colorlinks=true,       % false: boxed links; true: colored links
   linkcolor=blue,        % color of internal links
   citecolor=red,         % color of links to bibliography
   filecolor=magenta,      % color of file links
   }

\newcommand{\pmet}{$\cancel p_T$}

\begin{document}

% preprint numbers
\begin{flushright}
%IISC/TH/15-34
\end{flushright}

% title page ...

\title{Extraction of Neutrino Yukawa Parameters from Displaced Vertices of Sneutrinos}    
\author[1,2]{Amit Chakraborty,}
\author[3,4]{Stefano Moretti,}
\author[3,4]{Claire H. Shepherd-Themistocleous}
\author[3,4]{and Harri Waltari}

\emailAdd{amit.c@srmap.edu.in}
\emailAdd{S.Moretti@soton.ac.uk}
\emailAdd{Claire.Shepherd@stfc.ac.uk}
\emailAdd{H.Waltari@soton.ac.uk} 

\affiliation[1]{Department of Physics, School of Engineering and Sciences,\\ 
SRM University AP, Amaravati, Mangalagiri 522240, India}

\affiliation[2]{Centre for High Energy Physics, \\ 
Indian Institute of Science, Bangalore 560012, India}

\affiliation[3]{Particle Physics Department, STFC Rutherford Appleton Laboratory, \\
Chilton, Didcot, Oxon OX11 0QX, UK}

\affiliation[4]{School of Physics and Astronomy, University of Southampton, \\
Highfield, Southampton SO17 1BJ, UK}

\date{\today}
%%%%%%%%%%%%%%%%%%%%%%%%%%%%%%%%%%%%%%

\abstract{We study displaced signatures of sneutrino pairs potentially emerging at the Large Hadron Collider (LHC) in a Next-to-Minimal Supersymmetric Standard Model supplemented with right-handed neutrinos triggering a Type-I seesaw mechanism. We show how such signatures can be established through a heavy Higgs portal, the sneutrinos then decaying to charged leptons and charginos giving rise to further leptons or hadrons. We finally illustrate how the Yukawa parameters of neutrinos can be extracted by measuring the lifetime of the sneutrino from the displaced vertices, thereby characterising the dynamics of the underlying mechanism of neutrino mass generation. We show our numerical results for the case of both the current and High-Luminosity LHC.}

%%%%%%%%%%%%%%%%%%%%%%%%%%%%%%%%%%%%%%
\maketitle

%%%%%%%%%%%%%%%%%%%%%%%%%%%%%%%%%%%%%%%%%%%%%%%
\newpage
\setcounter{footnote}{0}

%%%%%%%%%%%%%%%%%%%%%% Table of content %%%%%%%

%\hrule
%\tableofcontents
%\vskip 1.0cm
%\hrule
%%%%%%%%%%%%%%%%%%%%%%%%%%%%%%%%%%%%%%%%%%%%%%%

%===========================================================================
%===========================================================================
\section{Introduction}
\label{intro}

The Standard Model (SM) has survived nearly all experimental tests. Besides the strong evidence for Dark Matter (DM), 
the only unexplained experimental phenomenon is neutrino oscillations 
\cite{Athanassopoulos:1997pv,Fukuda:1998mi,Aguilar:2001ty,Ahn:2002up,Abe:2011sj,An:2012eh}, which in turn 
imply that neutrinos have a tiny, but non-zero, mass. The standard explanation is that neutrino 
masses are generated through a seesaw 
mechanism  \cite{Minkowski:1977sc,Konetschny:1977bn,Mohapatra:1979ia,Magg:1980ut,Schechter:1980gr,Foot:1988aq}, where 
the effective dimension five operator responsible for neutrino masses is suppressed by a heavy mass scale.

The range of possible seesaw scales varies  from eV-scale sterile neutrinos to those with masses of the order  
$10^{14}$~GeV. Taking type-I seesaw as an example,  the neutrino Yukawa couplings  are 
$\mathcal{O}(1)$ at the upper end while they are tiny at the lower end of the possible spectrum of their interaction strength. One interesting option is that the seesaw scale is around the Electro-Weak scale (EW), which requires the neutrino 
Yukawa couplings to be somewhat smaller than the electron Yukawa coupling. As the Right-Handed (RH) 
neutrinos are singlets under the SM gauge group, the only interactions they have are the Yukawa 
couplings, which being small can lead to Displaced Vertices (DVs) \cite{Basso:2008iv,Helo:2013esa,Izaguirre:2015pga,Accomando:2016rpc,Liu:2019ayx}.

Supersymmetry (SUSY) is a well-motivated framework for Beyond the SM (BSM) physics. SUSY is the 
only space-time symmetry that can be added to the Poincar\'e algebra \cite{Haag:1974qh} and it 
relates particles with different spins, specifically, bosons to fermions. This relation leads to the cancellation of quadratic divergences emerging in the calculation of the Higgs boson mass in the SM (the so-called hierarchy problem). 
Furthermore, SUSY may induce the convergence of the Electro-Magnetic (EM), weak and strong couplings at some high energy scale, unlike the SM, a precondition for a theory embedding unification of forces. Finally,  
if one removes the baryon and lepton number violating couplings by requiring $R$-parity, one can get as a by-product of SUSY a DM candidate in the form of the Lightest Supersymmetric Particle (LSP).

Needless to say then, in order to pursue BSM physics that addresses all the aforementioned SM flaws, SUSY is one of the possible paths to follow, so long that it embeds a mechanism for neutrino mass generation. In doing so, it is then necessary to surpass its minimal realisation and consider non-minimal ones \cite{Book}, wherein the gauge and/or Higgs structures are enlarged with respect to the case of the Minimal Supersymmetric Standard Model (MSSM). An attractive framework in this respect is the Next-to-MSSM (NMSSM), wherein a singlet Superfield, containing an extra singlet Higgs state and its SUSY counterpart, is added  to the MSSM particle content. This way, the so-called $\mu$-problem \cite{Ellwanger:2009dp} of the MSSM is overcome. If such a construct is supplemented with RH neutrinos and their SUSY counterparts, a viable model for neutrino mass generation based on a type-I seesaw is established. 

By adopting this theoretical framework, we will show that the heavy Higgs states belonging to it (both CP-even and -odd) can have significant  couplings to RH sneutrinos, 
 even in the alignment limit, as required by measurements of the SM-like Higgs boson discovered at the Large Hadron Collider (LHC) in 2012. Furthermore, 
given that  in this model RH (s)neutrinos and higgsinos get their masses through the same mechanism, one can expect the SUSY states to be rather degenerate, yet, suitable soft SUSY-breaking mass terms can render the RH sneutrinos somewhat  heavier than the higgsinos. In such a case a RH sneutrino can decay visibly through its Yukawa interactions to a charged lepton ($\ell = e, \mu$) and chargino or else a neutrino and neutralino. Therefore, a typical signal that may emerge at the LHC in
this theoretical scenario is heavy Higgs mediated production of a sneutrino pair eventually yielding  a di-lepton signature together with soft jets and missing transverse energy. As the seesaw mechanism has a source of lepton number violation, we get both opposite-sign and same-sign dileptons, the latter giving better discovery potential due to smaller backgrounds.
Remarkably, the aforementioned mass degeneracy may make the sneutrinos long lived, so that the visible tracks of this signature
may be displaced, which in turn implies a smaller background with respect to the one affecting similar prompt signatures \cite{Moretti:2019yln,Moretti:2020zbn}. 
Here we shall prove that this signature with two DVs can be extracted at the LHC and, moreover, we shall also show how  
the kinematics of the displaced (visible) tracks  could allow for a measurement of the (s)neutrino Yukawa couplings, thereby enabling one to probe the underpinning neutrino mass generation dynamics.

Our paper is organised as follows. In the next section we introduce our theoretical framework. In the following one we illustrate  the properties of the track displacements and how these can be related to the discussed Yukawa couplings. Then we perform our MC analysis aimed at extracting both the relevant signature and its underlying (s)neutrino mass parameters. We then conclude.

%%-------------------------------------------------------
%===========================================================================
\section{NMSSM with RH neutrinos}

We shall study the  NMSSM with RH neutrinos.
It is based on the following Superpotential \cite{Kitano:1999qb,Cerdeno:2008ep}

%\begin{multline}
$$
W=y^{u}_{ij}(Q_{i}\cdot H_{u})U^{c}_{j}-y^{d}_{ij}(Q_{i}\cdot H_{d})D^{c}_{j}-y^{\ell}_{ij}(L_{i}\cdot H_{d})E^{c}_{j}+y^{\nu}_{ij}(L_{i}\cdot H_{u})N^{c}_{i} + \lambda S(H_{u}\cdot H_{d})
$$
\begin{equation}\label{eq:superpotential}
\hspace*{-9.85cm}
+\frac{\lambda_{Ni}}{2}SN_{i}^{c}N_{i}^{c}+\frac{\kappa}{3} S^{3}.
\end{equation}
%\end{multline}

As mentioned, this model cures some problems of the MSSM, namely the $\mu$-term will be generated through the Vacuum Expectation Value (VEV) of the scalar component of the singlet Superfield $S$ and we also have a mechanism for neutrino mass generation. As the $\mu$-term should not be too far above the EW scale, the RH neutrino masses are at the EW scale too, hence the neutrino Yukawa couplings need to be very small, of the order $10^{-7}$.

We shall look at a scenario where we have light higgsinos and RH neutrinos being roughly degenerate with these. The soft SUSY-breaking masses should then make the RH sneutrinos heavier than the higgsinos and thus the decays $\tilde{N}\rightarrow \tilde{\chi}^{0}\nu,\tilde{\chi}^{\pm}\ell^{\mp}$ are kinematically open. The decay width will be given by the neutrino Yukawa couplings. As mentioned, they will be tiny, and thus may lead to DVs.

This model allows two important features: EW Symmetry Breaking (EWSB) generates both a lepton-number violating mass term for the RH sneutrinos and a coupling between Higgs states and the sneutrinos. The coupling in the alignment limit is (neglecting doublet-singlet mixing)
\begin{eqnarray}
C_{h\tilde{N}\tilde{N}} & = & \pm\frac{1}{2}\lambda\lambda_{N}v\sin 2\beta,\\
C_{H\tilde{N}\tilde{N}} & = & \pm\frac{1}{2}\lambda\lambda_{N}v\cos 2\beta,
\end{eqnarray}
where the upper (lower) sign is for CP-even (CP-odd) sneutrinos. If $\tan \beta > 1.5$, the heavy Higgs state has a stronger coupling to sneutrinos. If $\lambda$ and $\lambda_{N}$ are large, RH sneutrinos can be pair produced through the heavy Higgs portal 
and they can be detected through lepton-number violating signatures \cite{Moretti:2019yln,Moretti:2020zbn}. The singlet field is essential in achieving this as in the MSSM with RH neutrinos the Higgs-RH sneutrino couplings would not exist.
As intimated, our aim is to use DVs to both improve background rejection and allow for a quantitative estimate of the neutrino Yukawa couplings.

%===========================================================================
\section{From displacements to Yukawa couplings} \label{sec:yukawa1}

We shall assume that the only kinematically available decay channels for the sneutrino are $\tilde{N}\rightarrow \tilde{\chi}^{0}\nu,\tilde{\chi}^{\pm}\ell^{\mp}$ and that the neutralino and chargino are higgsino-like.

The sneutrino-lepton-chargino vertex factor is

\begin{equation}
\lambda_{\tilde{N}\ell^{+}\tilde{\chi}^{-}}=\frac{i}{\sqrt{2}}y^{\nu}_{ab}V_{12}\frac{1+\gamma_{5}}{2},
\end{equation}
where $a,b$ refer to the flavours of the charged lepton, sneutrino and $V_{12}$ tells the higgsino component of the lightest chargino. For CP-odd sneutrinos we only need the replacement $\frac{i}{\sqrt{2}}\rightarrow \frac{1}{\sqrt{2}}$. This leads to the partial width (neglecting the lepton mass)

\begin{equation}\label{eq:charginowidth}
\Gamma(\tilde{N}_{i}\rightarrow \ell^{\pm}_{j}\tilde{\chi}^{\mp})=\frac{(m_{\tilde{N}}^{2}-m_{\tilde{\chi}^{\pm}}^{2})^{2}}{16\pi m_{\tilde{N}}^{3}}|y^{\nu}_{ji}|^{2}|V_{12}|^{2}.
\end{equation}
 We shall assume $|V_{12}|=1$ in the following.

If we neglect the mixing between Left-Handed (LH) and RH neutrinos\footnote{This mixing introduces a vertex factor $\lambda_{N}N_{j5}$ times the RH neutrino component of the light eigenstates. The elements of the neutrino left-right mixing matrix are of the same order as the elements of $y^{\nu}$. As the singlino component $N_{j5}$ is not negligible, this correction to the vertex factor is numerically only an order of magnitude smaller than $|y^{\nu}N_{j3}|$ so this does introduce an $\mathcal{O}(10\%)$ correction to the partial width.\label{mixing}}, the sneutrino-neutrino-neutralino vertex factor is

\begin{equation}
\lambda_{\tilde{N}\nu\tilde{\chi}^{0}}=-\sum_{a}\frac{i}{\sqrt{2}}\left( N_{j3}^{*}P^{*}_{ab}y^{\nu}_{bc}\frac{1-\gamma_{5}}{2}+
N_{j3}P_{ab}y^{\nu *}_{bc}\frac{1+\gamma_{5}}{2}\right),
\end{equation}
where $P$ is the 
Pontecorvo-Maki-Nakagawa-Sakata (PMNS) matrix, $b$ and $c$ are the flavours of the neutrino and sneutrino, respectively,  while $N_{j3}$ gives the $\tilde{H}_{u}$ component of the neutralino $j$. For CP-odd sneutrinos we shall again replace $\frac{i}{\sqrt{2}}\rightarrow \frac{1}{\sqrt{2}}$ in the prefactor. This gives the partial width

\begin{equation}\label{eq:neutralinowidth}
\Gamma(\tilde{N}_{i}\rightarrow \nu\tilde{\chi}^{0}_{j})=\sum_{k}\frac{(m_{\tilde{N}}^{2}-m_{\tilde{\chi}^{0}}^{2})^{2}}{16\pi m_{\tilde{N}}^{3}}|y^{\nu}_{ki}|^{2}|N_{j3}|^{2}.
\end{equation}
Here $j$ can get values $1,2$ and $|N_{j3}|^{2}\simeq 1/2$.

From Equations (\ref{eq:charginowidth}) and (\ref{eq:neutralinowidth}) we see that the total decay width is proportional to $\sum_{i} |y^{\nu}_{ij}|^{2}$. Hence the measurement of the lifetime of the sneutrino will give an estimate of the sum of the Yukawa couplings, while the ratios $|y_{ik}/y_{jk}|^{2}$ are proportional to BR$(\tilde{N}_{k}\rightarrow \ell_{i}^{\pm}\tilde{\chi}^{\mp})/{\rm BR}(\tilde{N}_{k}\rightarrow \ell_{j}^{\pm}\tilde{\chi}^{\mp})$. Measuring the lifetime and the ratios of the Branching Ratios (BRs) would then give us the absolute values of the individual neutrino Yukawa couplings.

We shall assume that the chargino and neutralinos would have been observed already and their masses would be known reasonably well. From the decay mode $\tilde{N}\rightarrow \ell^{\pm}\tilde{\chi}^{\mp}$ and the subsequent chargino decay $\tilde{\chi}^{\pm}\rightarrow \tilde{\chi}^{0}+$ hadrons, the invariant mass of the visible decay products should have an endpoint at $m_{\tilde{N}}-m_{\tilde{\chi}^{0}}$ from which the sneutrino mass could be estimated. We also expect that the mass of the heavy Higgs bosons would be known from one of its fermionic decay modes.

In order to extract the lifetime of the sneutrino in its rest frame, we need to measure the position of the secondary vertex and the relativistic $\gamma$-factor of the sneutrino, which can be computed, \textit{e.g.}, through $\gamma_{\tilde{N}}=E_{\tilde{N}}/m_{\tilde{N}}$. This can be done on an event-by-event basis as follows.

In the Center-of-Mass (CM) frame energy-momentum conservation gives us

\begin{eqnarray}
E_{\tilde{N}}^{*} & = & \frac{m_{H}}{2},\\
p_{\tilde{N}}^{*} & = & \sqrt{\frac{m_{H}^{2}}{4}-m_{\tilde{N}}^{2}}.\label{eq:labmomentum}
\end{eqnarray}
To figure out what are the sneutrino momenta in the laboratory frame, we may use the following facts.

\begin{itemize}
\item The initial transverse momentum of the heavy Higgs boson is nearly zero as long as there are no hard prompt jets in the event.

\item The three-momentum of the sneutrino is directed from the primary vertex to the secondary one, since the sneutrino is a neutral particle and its trajectory will not be curved.

\end{itemize}

%--------------------------------------
\begin{figure}[htb]
\begin{center}
\includegraphics[width=\textwidth]{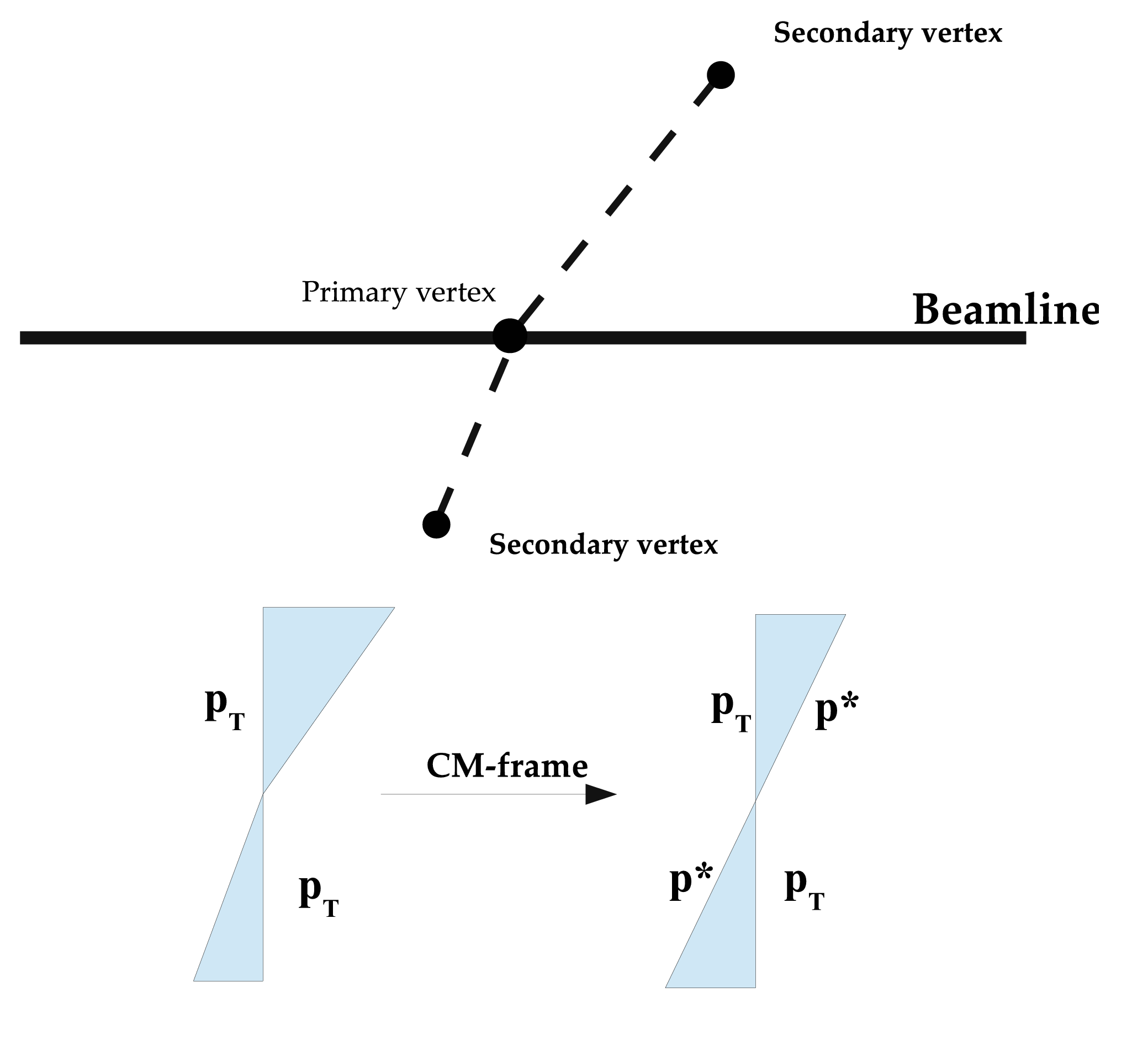}
\end{center}
\caption{The kinematics can be solved as shown. As the sneutrinos are neutral, their momentum vectors are aligned with the displacements of the secondary vertices from the primary vertex. We then construct the momenta in the lab frame by using these direction vectors and the fact that the transverse components are equal. Then we may boost to the center-of-mass frame, where we know the total momentum $p^{*}$ drom Eq. (\ref{eq:labmomentum}). \label{fig:kinematics}}

\end{figure}
%--------------------------------------

\begin{figure}[!htb]
\begin{center}
\includegraphics[width=0.5\textwidth]{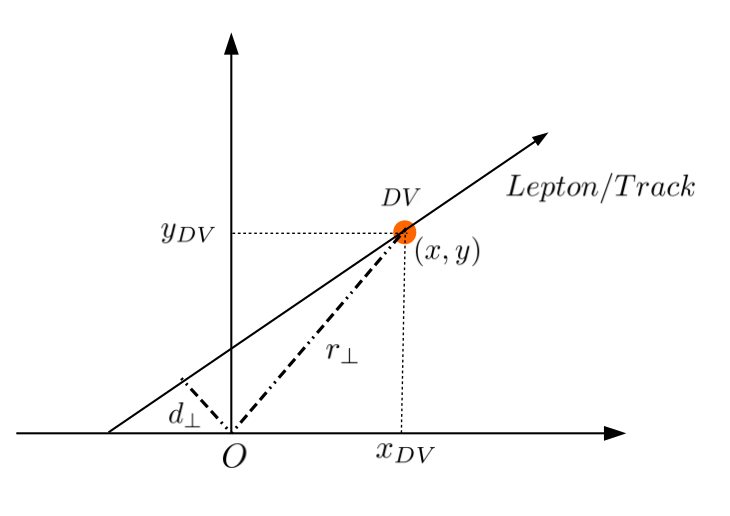}
\end{center}
\caption{Schematic view in the transverse plane of a long-lived particle decay.}
\label{fig:dv}
\end{figure}

We shall draw vectors from the primary vertex to the secondary vertices as shown in Figure \ref{fig:kinematics}. Next we shall scale them so that $|p_{T,\tilde{N}_{1}}|=|p_{T,\tilde{N}_{2}}|$ based on the small initial transverse momentum. Due to the difference of initial state gluon momenta, the final state will have a net momentum in the $z$-direction, but we may boost to the CM frame simply by adding equal vectors to both laboratory frame momenta. Once we are in the CM frame, we know $p_{\tilde{N}}^{*}$ from Equation (\ref{eq:labmomentum}) and hence may solve for $p_{T,\tilde{N}}$. This then allows us to compute

\begin{equation}
p_{\tilde{N}}=p_{T,\tilde{N}}\sqrt{1+\frac{d_{Z}^{2}}{d_{\perp}^{2}}},
\end{equation}
where $d_{z}$ and $d_{\perp}$ are the longitudinal and transverse displacements, respectively. In Figure \ref{fig:dv}, we show a 
representative image of the trajectory of a long-lived particle in a 2-dimensional plane (from \cite{Allanach:2016pam}).

%--------------------------------------
%--------------------------------------

We can then deduce $E_{\tilde{N}}=\sqrt{p_{\tilde{N}}^{2}+m_{\tilde{N}^{2}}}$ and from this $\gamma_{\tilde{N}}$. As the total displacement is
\begin{equation}
d=\gamma_{\tilde{N}} v \tau,
\end{equation}
where $v$ is the velocity and $\tau$ the lifetime of the sneutrino, we may solve for the lifetime

\begin{equation}
\tau =\frac{d}{\gamma_{\tilde{N}}\sqrt{1-\frac{1}{\gamma_{\tilde{N}}^{2}}}}.
\end{equation}

The lifetimes follow an exponential probability distribution
\begin{equation}
P(\tau)\propto e^{-\tau/\tau_{0}},
\end{equation}
where $\tau_{0}$ is the average lifetime. Plotting the lifetime distribution with a suitable binning with a logarithmic scale should then produce a straight line, whose slope gives the inverse of the average lifetime\footnote{We refer to \cite{Banerjee:2019ktv,Liu:2020vur} for elaborated discussions on determining the lifetime of long-lived particles at the LHC.}

When Next-to-Leading Order (NLO) and even higher order corrections are taken into account, the Higgs bosons will have a finite initial transverse momentum $p_{T}^{H}$. This will introduce a relative error in the measurement of the momenta, which is of the order of $p_{T}^{H}/p_{\tilde{N}}$. As we are interested in events that do not have a large longitudinal boost, \textit{i.e.}, all decay products are in the barrel region of the detector, we expect typically the $\gamma$-factor of the sneutrino to be rather small. In such a case the error on $\gamma_{\tilde{N}}$ will be of the order of $(p_{T}^{H}/m_{\tilde{N}})^{2}$. If we veto for hard jets a typical $p_{T}^{H}$ is about $10$~GeV \cite{Harlander:2014uea} and the mass of the sneutrino is $200$~GeV or more so the error from the finite initial transverse momentum on the lifetime measurement is very small.

%===========================================================================

%===========================================================================
%===========================================================================
%\section{Event selection and backgrounds}
\section{Collider analysis}

We shall look at the process $pp\rightarrow H,A \rightarrow \tilde{N}\tilde{N}$, where both sneutrinos 
decay visibly giving a charged lepton ($\ell = e, \mu$) and a charged higgsino. In this model the sneutrinos do not 
have a well-defined lepton number and the lepton number violating mass splitting is larger than 
the decay width, so there is a $50\%$ chance for each sneutrino to decay to either charge 
lepton. As the backgrounds for same-sign dileptons are smaller, we choose events with 
two same-sign leptons and veto for a third hard lepton with $p_{T}>15$~GeV.

If the higgsinos have a relatively smaller mass splitting, say close to $5-10$~GeV, then 
the decay width for the sneutrino to visible decays is small. Numerically the 
minimal width is close to $10^{-14}$~GeV leading to a mean decay length of a couple of 
millimeters. In contrast, in the region of phase space where the production of heavy 
Higgses is possible with a reasonable cross section and the decays are kinematically allowed, the 
decay widths are typically a few times $10^{-13}$~GeV. This implies that the mean decay lengths 
are around hundreds of micrometers, which leads to final states with DVs. 
Thus for a large fraction of the signal events we should have two DVs with 
charged leptons and soft charged tracks. 

In this work, we require both of the leptons ($\ell = e, \mu$)
to be displaced. Such a requirement means that the 
backgrounds come from processes involving either $b$-quarks or $c$-quarks. The same-sign requirement 
further reduces these as in the case of only two heavy-flavour quarks 
same-sign leptons are possible only in the case of flavour oscillations.

Our signal events will have missing transverse momentum in the form of the LSPs and, 
if the spectrum is not compressed, it is possible to require large $\slashed{p}_{T}$ with a 
rather good signal acceptance. Heavy flavour events rarely satisfy this requirement. In 
order to have neutrinos with significant transverse momentum, the quarks in the hard 
process must have large $p_T$ and the cross section falls down quickly with increasing 
$p_{T}$. Also, $t\overline{t}$ events with both the tops decaying hadronically 
can kinematically mimic the signal events.

%-----------------------------------------
\subsection{Event generation procedure}

For detailed analysis we choose the three Benchmark Points (BPs) given in Table \ref{tab:benchmark}. While they differ slightly in their mass spectrum, their main difference is in the sneutrino lifetimes.  BP-I represents a ``typical'' benchmark for EW scale seesaw with a decay width corresponding to a mean decay length around $1$~{\rm mm}. The second benchmark has a shorter lifetime with a mean decay length less than $0.3$~{\rm mm}, while the third one has a mean decay length around $2$~{\rm mm}, which is about as long as one can get without going to very compressed spectra. For such cases any leptons would be soft and triggering the event would be more difficult.
For interested readers, we refer \cite{Fukuda:2019kbp,Bhattacherjee:2020nno} for the recent proposals on these issues.  

Regarding the constraints on the spectrum, the higgsinos need to be heavier than about $160$~GeV with a mass splitting of $10$~GeV \cite{Sirunyan:2018iwl}. Since $m_{H}>2m_{\tilde{N}}>2m_{\tilde{H}}$, the heavy Higgses need to be beyond $400$~GeV. However, the production cross section falls off rather quickly beyond $500$~GeV. In this range $\tan \beta$ needs to be low to avoid the constraints from $H\rightarrow \tau^{+}\tau^{-}$ searches \cite{Sirunyan:2018zut,Aad:2020zxo}. For our BPs we have $2<\tan\beta < 3$ as this both evades the experimental constraints and gives a large BR($H\rightarrow \tilde{N}\tilde{N}$).

We simulate 100,000 signal events and $\mathcal O(10^{7})$ events 
each for the $t\bar{t}$ and $b\bar{b}$ backgrounds using {\tt MadGraph5 v2.6.6} \cite{Alwall:2011uj} 
at LO. Parton showering and hadronisation are modelled through 
{\tt Pythia v8.2} \cite{Sjostrand:2014zea} and fast detector simulation is obtained by 
{\tt Delphes v3.3.3} \cite{deFavereau:2013fsa} with the ATLAS card. We use a modified version of the default ATLAS card to implement the impact parameter smearing effects. 
The event rates are then corrected to NLO accuracy 
with $k$-factor of 2 for the signal \cite{Spira:1993bb,Spira:1995rr,Muhlleitner:2006wx} and to 
next-to-next-to-leading order (NNLO) accuracy with $k$-factor of 1.8 for the two dominant 
backgrounds \cite{Czakon:2011xx, Aliev:2010zk, Catani:2020kkl}\footnote{Note that, as pointed out in 
Ref.\cite{Catani:2020kkl}, at high transverse momenta of the bottom quarks, large 
logarithmic terms of the form $\ln(\frac{p^b_T}{m_b}) $ become important and need to be resummed properly while 
estimating the NNLO cross section for the $b\bar{b}$ process. For our study, bottom quarks are pair produced 
with a minimum $p^b_T$ = 200 GeV, so we make a conservative choice of $k$-factor = 1.8.}. Further, in 
order to generate events efficiently, we demand that the top quarks are decaying hadronically and 
the $b$-hadrons, obtained after parton shower and 
hadronisation of the $b$-quarks, are decaying through leptonic final states. We generate two sets of 
$b\bar b$ samples by varying the $p_T$ of the bottom quarks at the generation level. We find that 
the one with generation level cut $p^{b}_{T, {\rm min}}$ = 200 GeV has better sensitivity. In Table \ref{tab:events}, we show the details of event generation of individual signal and background events.

%----------------------------------------------------------
\begin{table}[!htb]
\begin{center}
%\scalebox{0.8}{
\begin{tabular}{ |c|c|c|c| }
\hline
Observable & BP-I & BP-II & BP-III \\
\hline
\hline
Lightest Higgs mass & 125.0 & 125.2  & 125.6 \\
\hline
2nd Higgs mass & 338.2  & 322.1 & 370.4\\
\hline
3rd Higgs mass & 462.2 & 484.0 & 483.6 \\
\hline
Lightest Pseudoscalar Higgs mass & 259.0  & 256.8 & 261.7\\
\hline
2nd Pseudoscalar Higgs mass  & 446.9 & 470.2 & 468.1 \\
\hline
\hline
Lightest Sneutrino mass & 219.4  & 219.9 & 228.4  \\
\hline
Lightest CP-odd Sneutrino mass & 220.0 & 219.8   & 229.2  \\
\hline
\hline
Lightest Chargino mass  & 185.8  & 177.6 & 205.1 \\
\hline
Lightest Neutralino mass  & 177.3 & 168.2 &  196.0 \\
\hline
Next-to Lightest Neutralino mass & 200.3  & 193.2 &  218.6 \\
\hline
\hline
BR($h_3 \to \tilde{\nu_1} \tilde{\nu_1}$) (in \%) & 5.3  & 4.9  & 4.9 \\
\hline
BR($A_3 \to \tilde{\nu_1} \tilde{\nu_1}^\prime$) (in \%) & 1.2 & 1.3 & 1.6 \\
\hline
BR($\tilde{\nu_1} \to \ell \tilde{\chi^{\pm}_1}$) (in \%) & 48.2 & 48.6 & 45.3  \\
\hline
BR($\tilde{\nu_1} \to \nu \tilde{\chi^{0}_1}$) (in \%)& 51.8 & 51.4 & 54.7 \\
\hline
$\Gamma(\tilde{\nu_1})$ (GeV) & $1.6 \times 10^{-13}$  & $8.5 \times 10^{-13}$ & $9 \times 10^{-14}$ \\
\hline
\end{tabular}
%}
\caption{Details of the BPs (all the masses are in GeV). The leptonic BRs 
include electron, muons and taus.}
\label{tab:benchmark}
\end{center}
\end{table}
%----------------------------------------------------------

%----------------------------------------------------------
\begin{table}[!htb]
\begin{center}
%\scalebox{0.8}{
\begin{tabular}{ |c|c|c|c| }
\hline
Process & Cross section (pb) & Events generated & Event weight factor \\
\hline
Signal (BP-I) & 0.0666 & 100,000 & 0.09 \\
\hline
Signal (BP-II) & 0.0558 & 100,000 & 0.08 \\
\hline
Signal (BP-III) & 0.0508  & 100,000 & 0.07 \\
\hline
$t\bar{t}$ (hadronic)& 369.0 & 10,000,000 & 5.1 \\ 
\hline
$b\bar b$ ($p^{b}_{T, {\rm min}}$ = 30 GeV)  & 1183654.3  & 5,000,000 & 32432.1  \\
\hline
$b\bar b$ ($p^{b}_{T, {\rm min}}$ = 200 GeV)  & 378.0 & 10,000,000 & 5.2  \\
\hline
\end{tabular}
%}
\caption{Event simulation details at $\sqrt s = 13 ~{\rm TeV}$ and $\mathcal L = 137~{\rm fb}^{-1}$.}
\label{tab:events}
\end{center}
\end{table}
%----------------------------------------------------------

%-----------------------------------------------
\subsection{Definition of displaced objects} 

We now provide the details of the observables used to probe the displaced signal events. 
\begin{itemize} 
\item \underline{Displaced leptons}: The isolated leptons ($\ell = e, \mu$) with $p_{T}(\ell)>10$~GeV and $|\eta(\ell)| < 2.5$ must 
satisfy $|{d_\perp}|> 0.2~{\rm mm}$ \cite{CMS:2014hka, Aad:2019tcc} where $d_{\perp}$ is the transverse impact 
parameter relative to the primary vertex. The lepton isolation is achieved by demanding the angular separation between the lepton and jets, $\Delta R (\ell, jet)$, should be greater than 0.4. Additionally, we demand that the leptons carry at least 80\% (90\%) of the transverse momentum within a cone of radius R = 0.5 in case of a muon (electron). Note that, we have used a modified version of the default ATLAS card available within {\tt Delphes} to implement the impact parameter smearing effects and obtain the displaced leptons.

\item \underline{DVs using displaced tracks}: 

The tracks used in the DV reconstruction must satisfy the following requirements: $p_{T}>1$~GeV and $|{d_\perp}| > 2~{\rm mm}$. 
Further, the significance of $d_\perp$ with respect to the beam axis ({\it i.e.}, $|d_\perp|$ divided by its uncertainty $\sigma_{d_\perp}$) 
should be at least 4 \cite{Sirunyan:2018pwn,Aad:2019kiz,Aad:2019tcc}.  The final requirement on 
track $\frac{|d_\perp|}{\sigma_{d_\perp}}$ improves the identification of displaced tracks associated to 
the DVs. 

We collect the displaced tracks and construct the DV. To combine the tracks, we 
use the truth information of the vertices ({\it i.e.}, vertex position) obtained from the detector emulator and merge those tracks 
if $|\Delta X (t_i, t_j)| < 0.001$, $|\Delta Y (t_i, t_j)| < 0.001$ and $|\Delta Z (t_i, t_j)| < 0.001$, where 
$t_i$ denotes the $i$-th displaced track. The invariant mass of the DVs are calculated using the 
summed 4-momenta of the associated tracks, {\it i.e.}, $m^2_{\rm DV} = {(\sum E_i)}^2 - {(\sum {\vec p}_i)}^2$, where 
the sum runs over the tracks associated to the DV \cite{Aad:2019kiz}.

\item \underline{Jets and displaced jets}: The jets are constructed from 
calorimeter tower elements using {\tt Fastjet v3.3.2} \cite{Cacciari:2011ma} and the anti-$k_T$ jet 
clustering algorithm \cite{ Cacciari:2008gp} with jet radius $R = 0.5$. We demand that the jets must satisfy $p_{T}>20$~GeV and 
$|\eta| < 3.0$. For signal events, 
hadronic decay of the charginos leads to displaced hadronic final states. Note that long-lived hadrons 
({\it e.g.}, $b$-hadrons, $c$-hadrons) as well as soft particles coming from the prompt decay of the displaced 
hard processes are also present in the events. So, we calculate the angular separation $\Delta R$ between the jet and the displaced tracks then demand that the displaced jet should have at least 2 or more displaced tracks satisfying $\Delta R (j,t) < 0.4$. The displaced jets are constructed following the Refs. \cite{Nemevsek:2018bbt, LLPtalk}. We check that for signal events, the final state stable objects are not energetic enough to pass the jet $p_T$ threshold and provide significant separation from the background events, therefore, we do not consider the displaced jets for further analysis.

\end{itemize}

%-----------------------------------------------
\subsection{Distribution of different observables} 

Here we show the distribution of several kinematic variables relevant for the collider analysis. All the histograms are drawn for events which satisfy the basic selections on the 
leptons and jets discussed in the previous section. Distributions are scaled for 
$\mathcal L = 137 ~{\rm fb}^{-1}$ of integrated luminosity at the $\sqrt s = 13$ TeV run of the LHC.

%-------------------------------
\begin{figure}[htb!]
\centering
\includegraphics[scale=0.3]{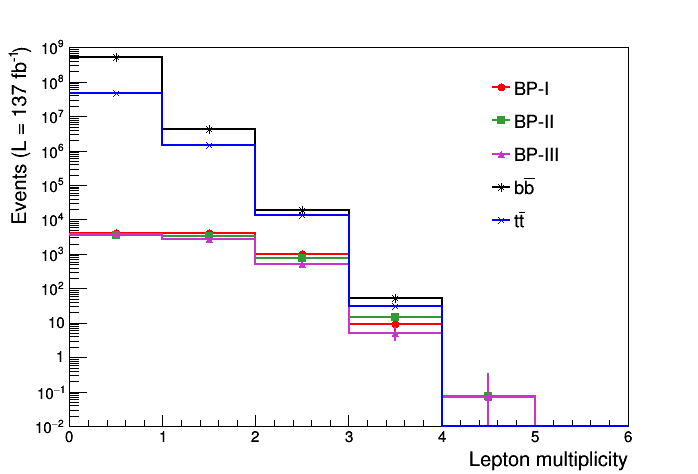}
\includegraphics[scale=0.3]{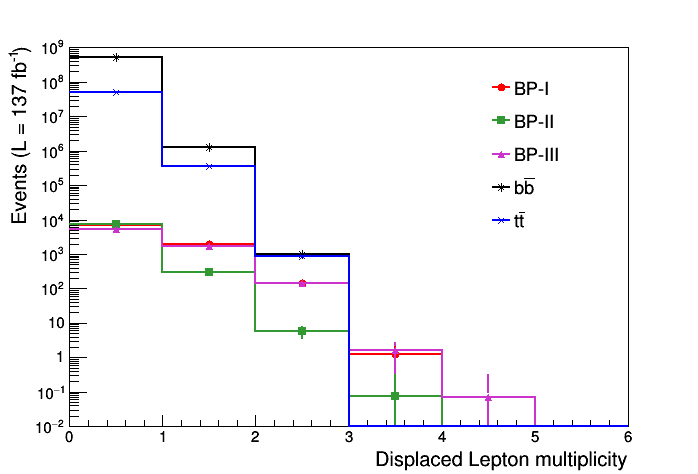}
\caption{Lepton multiplicity: all (left) and displaced (right).} 
\label{fig:nlep}
\end{figure}
%-------------------------------

%-------------------------------
\begin{figure}[htb!]
\centering
\includegraphics[scale=0.3]{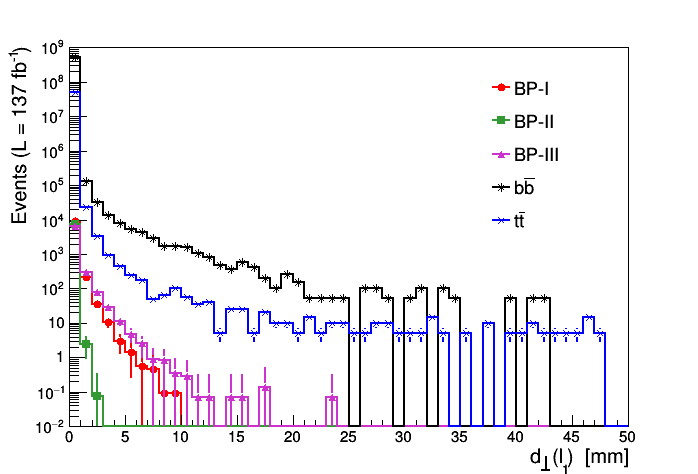}
\includegraphics[scale=0.3]{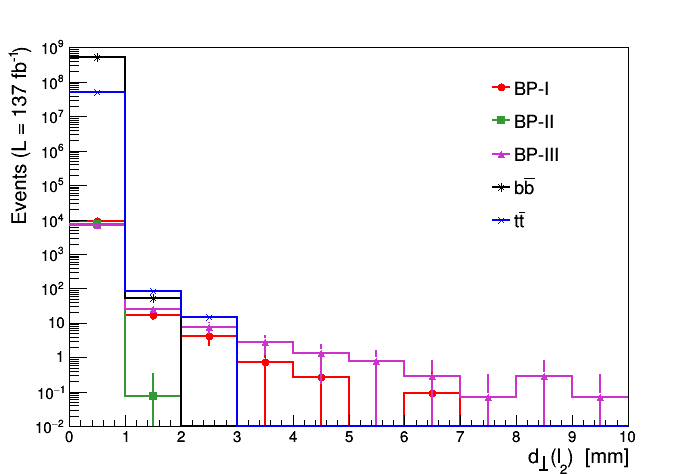}
\caption{Transverse impact parameter ($d_{\perp}$) of leading (left) and 
sub-leading (right) lepton. }
\label{fig:dxy}
\end{figure}
%-------------------------------

%-------------------------------
\begin{figure}[htb!]
\centering
\includegraphics[scale=0.3]{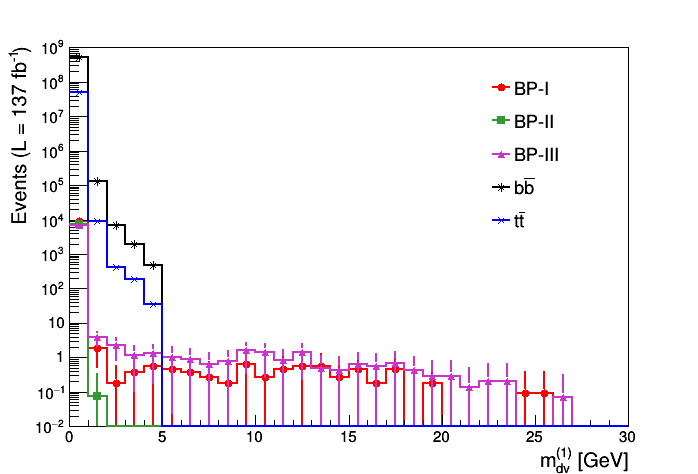}
\caption{Leading DV mass.}
\label{fig:mdv}
\end{figure}
%-------------------------------

In the left panel of Figure \ref{fig:nlep}, we show the multiplicity of isolated leptons present in the signal and 
background events. The right panel displays the same distribution but for the isolated displaced leptons with 
minimum transverse impact parameter $d_{\perp} > 0.2 ~{\rm mm}$. The $d_{\perp}$ distribution of the leading (in $p_T$) 
two leptons are shown in Figure \ref{fig:dxy}. Significant overlap is overserved expecially at  small 
values of $d_{\perp}$. As mentioned in the previous section, we construct the DVs using 
the displaced tracks and calculate the mass of the DV. In Figure \ref{fig:mdv}, we plot the 
mass of the leading (in mass) DV. It is evident that the presence of high $p_T$ displaced tracks, associated with 
the displaced leptons originating from the sneutrino decay, results in DVs with relatively larger 
masses. Therefore, we can control the background events by selecting events with mass greater than 5 GeV or so.
Also, the distribution has a kinematical endpoint at $m_{\tilde{N}}-m_{\tilde{\chi}^{\pm}}$ so this can be used to 
estimate the sneutrino mass once the chargino mass is known.

%-----------------------------------------------
\subsection{Event selection and signal significance} 

After looking at the distributions of several kinematic observables, we find the 
following selection cuts optimised for our process of interest. The cuts are 
as follows.

\begin{itemize}
\item C1: Two same-sign same-flavour leptons (electron or muons) satisfying basic lepton selection criteria. 
\item C2: The leading (in $p_T$) lepton must satisfy $p_{T}(\ell_{1})>25$~GeV.  
\item C3: The subleading (in $p_T$) lepton must satisfy $p_{T}(\ell_{2})>15$~GeV. 
\item C4: Veto on a third lepton with $p_{T}(\ell_{3})>15$~GeV. 
\item C5: For opposite sign same flavour leptons, veto  di-lepton invariant masses around the $Z$ mass 
{\it i.e.}, $m_{\ell^{\pm}\ell^{\mp}} \ne [80,100]$ GeV.
\item C6: Select events with \pmet\ $>$ 30 GeV.  
\item C7: Both  leptons have a transverse impact parameter $d_{\perp} > 0.2~{\rm mm}$. 
\end{itemize}
%----------------------------------------------------------

Even though our main backgrounds are from heavy-flavour jets, in this study we did not want to impose a $b$-veto so that we would be free of uncertainties related to the $b$-tagging when we show the viability of our approach.
As the displacement of a secondary vertex is a key input for $b$-tagging algorithms, the background displacement distributions would be affected by imposing a $b$-veto. Also the acceptance of the signal events is uncertain.
If experimental collaborations were to do this type of an analysis, they may be able to further improve the background rejection with the use of a $b$-tagger.

The complete cutflow  is presented in Table \ref{tab:cutflow}. The signal 
significance ($\frac{S}{\sqrt{(S+B)}}$), calculated at $\sqrt s = 13 ~{\rm TeV}$ 
and $\mathcal L = 137~{\rm fb}^{-1}$, is shown in Table \ref{tab:signi}. We may
see that the displacement requirement rejects almost all of the signal of BP-II.
For such a case the prompt signature can still be visible --- this is actually BP3 of
\cite{Moretti:2019yln} for which a cut-based analysis gave a $\sim 3\sigma$ excess.
From Table \ref{tab:cutflow} it is interesting to note that even though the cuts C4 and C5 
do not reduce the dominant backgrounds, however, they are important to reduce the sub-dominant 
backgrouds like WZ, ZZ, tW, tZ and $t\bar tZ$ processes where both prompt and displaced leptons can be 
present.

%----------------------------------------------------------
\begin{table}[!htb]
\begin{center}
%\scalebox{0.8}{
\begin{tabular}{ |c|c|c|c|c|c|c| }
\hline
 & BP-I & BP-II & BP-III & $t\bar{t}$ & $b\bar b$ (30 GeV) & $b\bar b$ (200 GeV)    \\
\hline
C1 & 420.4 & 354.9  & 160.6 & 2637.8 & 2.7 $\times 10^7$ & 440.2 \\ 
\hline
C2 & 354.7 & 335.8 & 71.9 & 857.0 & 2.4 $\times 10^6$ & 362.5 \\ 
\hline
C3 & 315.3  & 309.1 & 57.6 & 384.9 & 1.2 $\times 10^6$ & 176.1 \\ 
\hline
C4 & 314.7  & 307.8  & 56.9 & 384.9 & 1.2 $\times 10^6$ & 176.1 \\ 
\hline
C5 & 314.7 & 307.3 & 56.9 & 384.9 & 1.2 $\times 10^6$ & 176.1 \\ 
\hline
C6 & 265.8 & 270.4 & 49.6 & 123.2 & 32432.1 & 150.2 \\ 
\hline
C7 & 35.9 & 1.7 & 13.5 & 5.1 & 0 & 5.2 \\ 
\hline
\end{tabular}
%}
\caption{Cutflow table ($\sqrt s = 13 ~{\rm TeV}$ and $\mathcal L = 137~{\rm fb}^{-1}$).}
\label{tab:cutflow}
\end{center}
\end{table}
%----------------------------------------------------------

%----------------------------------------------------------
\begin{table}[!htb]
\begin{center}
%\scalebox{0.8}{
\begin{tabular}{ |c|c|c|c|c|c| }
\hline
 & Signal (S) & B ($t\bar{t}$) & B ($b\bar b$) & B (total) & Significance = $\frac{S}{\sqrt{(S+B)}}$   \\
\hline
BP-I & 35.9  & 5.1 & 5.2 & 10.3 & 5.3 \\ 
\hline
BP-II & 1.7  & 5.1 & 5.2 & 10.3 & 0.5 \\ 
\hline
BP-III & 13.5  & 5.1 & 5.2 & 10.3 & 2.8 \\ 
\hline
\end{tabular}
%}
\caption{Signal significances estimated at $\sqrt s = 13 ~{\rm TeV}$ and $\mathcal L = 137~{\rm fb}^{-1}$.\label{table:backgrounds}}
\label{tab:signi}
\end{center}
\end{table}
%----------------------------------------------------------

\begin{figure}[htb!]
\centering
\includegraphics[scale=0.3]{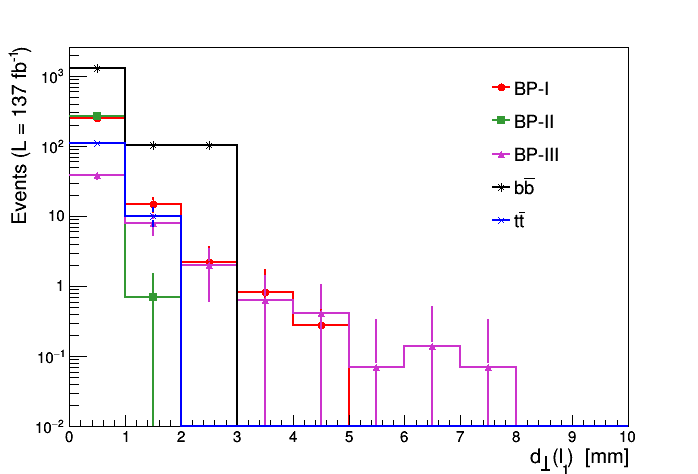}
\includegraphics[scale=0.3]{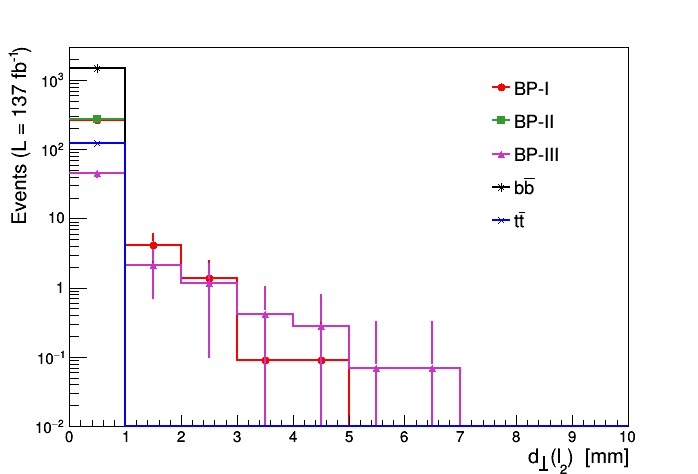}
\includegraphics[scale=0.3]{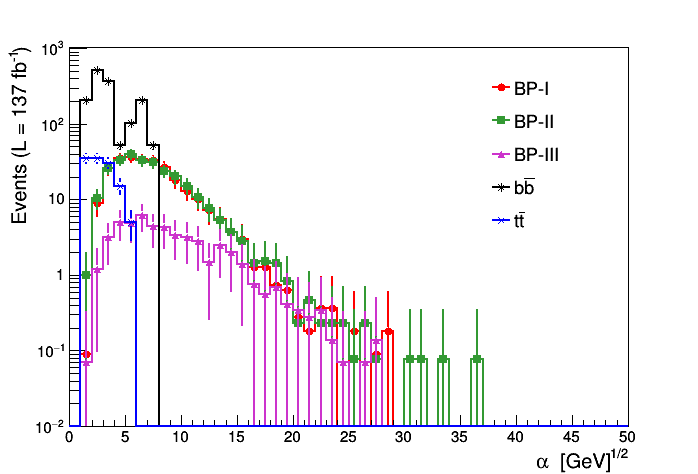}
\caption{Distribution of the transverse impact parameter ($d_{\perp}$) of the  leading (top-left) and
sub-leading (top-right) lepton. Bottom: The distribution of $\alpha = \frac{\cancel{p_T}}{\sqrt{H_T}}$. 
All the figures are drawn {for events satisfying cuts C1-C6}. }
\label{fig:aftercuts}
\end{figure}

Before we proceed to extract neutrino Yukawa couplings, we discuss a few 
kinematic observables which can be used, in addition to cuts C1-C6, to reduce the 
SM backgrounds. For example, we can optimise $d_\perp$ for both the 
leading two leptons, especially the second lepton which, for signal events, 
has larger $p_T$ and longer decay lengths (see plots in the 
upper panel of Figure \ref{fig:aftercuts}). Another important quantity is the ratio of 
the missing transverse momentum and scalar $H_T$, defined as 
$\alpha = \frac{\cancel{p_T}}{\sqrt{H_T}}$. Signal events have large missing transverse 
energy and, therefore, relatively larger values of $\alpha$. These additional observables 
provide us with an extra handle to minimise the SM backgrounds and improve the sensitivity 
to sneutrino events.

%-------------------------------
%-------------------------------

%===========================================================================
\section{Extraction of Yukawa couplings}

In order to extract the Yukawa couplings, we need to know the spectrum. The heavy Higgs masses can be measured reasonably accurately from their fermionic decay channels. The higgsinos would also be discovered from other searches and those would give an estimate for the masses of the neutralinos and charginos. The sneutrino mass can be estimated by measuring the invariant mass of the DV related to the sneutrino decay to a lepton and a chargino. The invariant mass distribution has an endpoint at $m_{\tilde{N}}-m_{\tilde{\chi}^{\pm}}$ (see Figure \ref{fig:mdv}), which will give us the sneutrino mass once we know the chargino mass.

We assume that we are able to get an essentially background free sample (say, over $95\%$ purity) using the cuts given in the previous section (both those given in Table \ref{tab:cutflow} and discussed at the end of the section) and the usage of $b$-tagging. We then do an event-by-event correction to the displacements described in section \ref{sec:yukawa1} to get the actual lifetimes. As both the CP-even and CP-odd Higgs bosons contribute to the signature, we need to pick which mass we use in the boost correction. As discussed in \cite{Moretti:2019yln}, the CP-even Higgs state  usually has the largest BR to sneutrinos, although also the amount of available phase space has an impact. If the CP properties of the two heavy Higgses have been measured, the CP-even mass would give the better estimate, otherwise it would be a reasonable choice to pick the heavier one due to the larger phase space. This ambiguity leads to a systematic error in the measurement of the Yukawa couplings, which is of the order of $10\%$ estimated by studying the impact of choosing the other Higgs mass to the lifetime distribution.

The background originating from heavy flavour hadrons that survives the cuts is heavily boosted (with the $p_{T}$ of a typical $b$-jet being $200$~GeV or more) leading to  an average displacement greater than $5$~mm. As the boost correction assumes the particle to be heavy, the lifetime of heavy flavour hadrons will be overestimated and will on average correspond to mean decay lengths around $5$~mm, although the exponential distribution of course gives events at all displacements. We shall use a binning of $0.1$~mm. As the total number of background events at the  HL-LHC is expected to be around $200$ (scaled from Table \ref{table:backgrounds} to $3000$~fb$^{-1}$) and these are distributed into some $100$ bins or more, there will not be many background events in a single bin. The background distribution may be estimated, \textit{e.g.}, by looking at events with small vertex mass and $\slashed{p}_{T}/\sqrt{H}_{T}$ and the subtract it from the distribution.

The signal region includes a cut on the transverse distance of the leptons requiring $d_{\perp}> 0.2$~mm. The distribution of lifetimes close to the cut will be modified and hence we put a lower bound on the lifetime when fitting. We fit the exponential to the distribution in the interval $0.5$~mm$ <c\tau<5$~mm. This leads to rather robust results if the true lifetime is around $c\tau \simeq 1$~mm but,  if the lifetime is shorter, the number of events in the fit region becomes so low that the statistical error increases. In such a case it might be reasonable just to use this method to find an upper limit for the lifetime and derive a lower limit for the neutrino Yukawa couplings. An upper limit for the Yukawa couplings can be obtained from the constraint on the sum of neutrino masses. That can be expressed as $\sum m_{\nu}=\mathrm{Tr}(m^{\nu})=\sum_{i,j}|y^{\nu}_{ij}|^{2}v^{2}\sin^{2}\beta/2m_{N_{j}}$.

Once we had studied the fitting method with one BP, we tried to analyse other BPs blindly so that one of the authors generated events and another one then analysed them,  the result was then compared to the unknown input. As long as the decay lenghts were at the mm scale, a reasonable agreement was achieved.
\begin{figure}
\begin{center}
\includegraphics[width=0.75\textwidth]{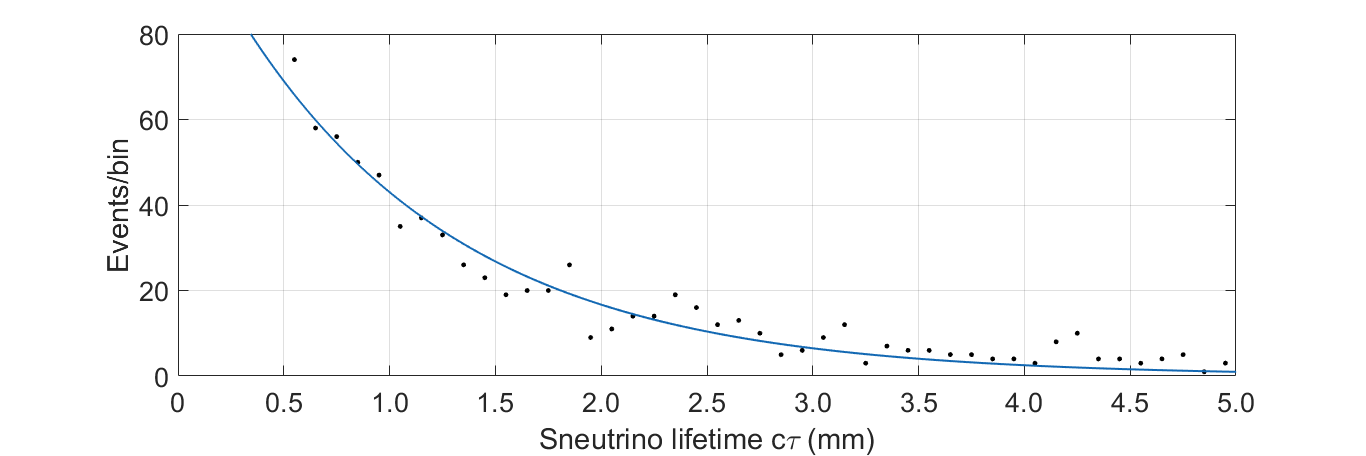}
\end{center}
\caption{The fit of an exponential to the lifetime distribution of the sneutrino for BP-I. The amount of data corresponds to $3000$~fb$^{-1}$ at $\sqrt{s}=14$~TeV.\label{fig:sneutrinofit}}
\end{figure}
We show in Figure \ref{fig:sneutrinofit} a fit of the sneutrino lifetime for BP-I. We fit a simple exponential to the lifetime distribution. The fit with a number of events corresponding to HL-LHC energy ($\sqrt s =14$ TeV) and 
luminosity ($3000$~fb$^{-1}$) gives us $\sum_{i}|y^{\nu}_{i1}|^{2}=(2.62 \pm 0.13 \pm 0.26) \times 10^{-13}$, where the first error is statistical and the second one the theory error (assumend to be $10\%$) from the ambiguity in choosing the CP-even or CP-odd Higgs as discussed above. On top of these there will be experimental systematic errors, which are mostly related to measuring the primary and secondary vertices\footnote{As this is a simple lifetime measurement, many typical sources for systematic errors, like parton distribution functions, do not matter.}.  The true value is $2.28\times 10^{-13}$, which slightly more than one standard deviation off.
For BP-II the event sample is so small that the Yukawa couplings cannot be reliably estimated. The best fit value would have been $3.57\times 10^{-13}$ but, as the actual average decay length is shorter than the lower end of our fitting region, this estimate is based only on few events and individual outliers can significantly change the result. The estimate was off by a factor of two. With such a short decay length it makes sense only to give a lower bound on the Yukawa couplings.
We give the results for our three BPs in Table \ref{tb:yukawas}. When doing the estimate, we added $10\%$ to the invisible decay width of Equation (\ref{eq:neutralinowidth}) due to the mixing effect of LH and RH neutrinos (see footnote on page \pageref{mixing}).

\begin{table}
\begin{center}
\begin{tabular}{|l|c|c|c|}
\hline
 & BP-I & BP-II & BP-III\\
\hline
Measured $\sum |y^{\nu}_{1i}|^{2}$ & $2.62\pm 0.13\pm 0.26$ & $>2.94$ & $3.73\pm 0.67 \pm 0.37$\\
Actual $\sum |y^{\nu}_{1i}|^{2}$ & $2.28$ & $7.81$ & $2.40$\\
\hline
\end{tabular}
\end{center}
\caption{The sums of neutrino Yukawa couplings from displacements and the actual input values in units of $10^{-13}$. The first uncertainty is statistical and the second one is a $10\%$ theoretical uncertainty related to the ambiguity between the CP-even/odd Higgs states as discussed in the text. For BP-II the decay length was so short that we only quote a lower limit for the Yukawa couplings.\label{tb:yukawas}}
\end{table}

From this value we can then determine the absolute values of individual Yukawa couplings as the number of events with a given lepton flavour, which will be proportional to $\epsilon_{i} |y^{\nu}_{i1}|^{2}$, where $\epsilon_{i}$ is the efficiency of identifying the lepton of flavour $i$. Since the error on $|y^{\nu}|^{2}$ will often be below $20\%$ with high enough statistics, the error on the Yukawa couplings themselves will be around $10\%$. Hence even this rather simple fitting method can give a reasonably good estimate of the Yukawa couplings.
%===========================================================================
\section{Conclusions and outlook}

We have studied sneutrino pair production via the heavy Higgs portal in the NMSSM with RH neutrinos. In the model both higgsinos and RH neutrinos get their masses through the singlet VEV, so we expect them to be at the EW  scale. If RH sneutrinos are heavier than higgsinos, they can decay to both a charged lepton and a chargino or a neutrino and a neutralino, the former one giving a visible signature at colliders.

Since the RH sneutrinos are gauge singlets, their decay modes are dictated by the neutrino Yukawa couplings, which are tiny in an EW scale seesaw model. The smallness of the decay width together with the lepton number violation arising from the RH (s)neutrino mass term leads to a signature with same-sign dileptons emerging from displaced vertices together with missing transverse momentum. The SM backgrounds for such a signal topology are therefore low.

We have indeed performed an alaysis proving that the study of the emerging signatures with DVs can lead to the extraction of the neutrino Yukawa couplings, which salient features are as follows. 

\begin{itemize}
\item We searched for displaced leptons and jets, with the leptons  originating from 
sneutrino decays. Several BPs of the aforementioned BSM scenarios were introduced, with varied displacement lengths, and studied by MC analysis assuming a  simple cut-based analysis providing  a good signal-to-backgroud  ratio already at the end of the 13 TeV run of the LHC. 
\item The kinematic distribution of the displaced vertex mass gives an end point which can be used to 
estimate the sneutrino mass once the chargino mass is known. 
\item Additional variables can make the regions of phase-space effectively background-free, thereby implying 
better handle on the extraction of the intervening Yukawa couplings. 
\item
The decay width of the RH sneutrinos are dictated by the sum of the squares of the Yukawa couplings so a measurement of the sneutrino lifetime would allow one to measure the Yukawa couplings. The individual Yukawa couplings can then be extracted from the sneutrino BRs to different lepton flavours, once corrected with identification efficiencies. 
\item We ultimately showed that, for lifetimes corresponding to mm scale displacements,  lifetimes can be measured with reasonable accuracy, which can then give the absolute values of the Yukawa couplings with a $10\%$ accuracy so long that sufficiently large data samples can be accrued for the signal, like, {\it e.g.}, at the HL-LHC.
\end{itemize}

In short, such an approach can lead to the extraction of the underlying neutrino dynamics parameters  from the study of sneutrinos signatures at hadronic colliders, albeit under a specific SUSY paradigm. The results obtained here from events with DVs therefore complement those obtained in Refs.~\cite{Moretti:2019yln,Moretti:2020zbn} for the case of prompt signatures.  
%===========================================================================
\section*{Acknowledgments}
SM, CHS-T and HW are   financed  in  part  through the  NExT  Institute. SM is also funded  by  the  STFC  consolidated  Grant  No.   ST/L000296/1. HW  acknowledges  financial  support from the Magnus Ehrnrooth Foundation, the Finnish Academy of Sciences and Letters and STFC Rutherford International Fellowship (funded through the MSCA-COFUND-FP Grant No.  665593). 
The work of AC is funded by the Department of Science and Technology, Government of India, under Grant No. IFA18-PH 224 (INSPIRE Faculty Award). Some of us also acknowledge the use of the IRIDIS High Performance Computing Facility and associated support services at the University of Southampton.
We thank Nishita Desai, Terhi J\"arvinen, Emmanuel Olaiya, Ian Tomalin and Jose Zurita for useful discussions.

%===========================================================================

%===========================================================================
%%%%%%%%%%%%%%%%%%%%%%%%%%%%%%%%%%%%%%%%%%%%%%%%%%%%%%%%%%%%%%%%%%%%%%%%%%%%%%%%%%%%%%%%%%%%%%
\end{document}